\documentclass{article}
\usepackage{graphics} 
\begin{document}
\begin{center}
{\bf{LOW ENERGY PION-PION ELASTIC SCATTERING IN SAKAI-SUGIMOTO MODEL}}

\vspace{0.5cm}

R.Parthasarathy {\footnote{e-mail address:sarathy@cmi.ac.in}} \\
Chennai Mathematical Institute \\
Plot.H1, SIPCOT IT Park, Padur Post \\
Siruseri 603103, India. \\

\vspace{0.3cm}

and \\

\vspace{0.3cm}

K.S.Viswanathan{\footnote{e-mail address:kviswana@sfu.ca}} \\
Department of Physics and IRMACS \\
Simon Fraser University \\
Burnaby, Canada V5A1S6.\\
\end{center}

\vspace{1.5cm}

{\noindent{\it{Abstract}}}

\vspace{0.5cm}

We have considered the holographic large $N_c$ QCD model proposed by Sakai and 
Sugimoto and evaluated the non-Abelian DBI-action on the $D8$-brane upto $(\alpha')^4$ 
terms. Restricting to the pion sector, these corrections give rise to four derivative 
contact terms for the pion field. We derive the Weinberg's phenemenological 
lagragian. The coefficients of the four derivative terms are determined in terms of 
$g^2_{YM}$. The low energy pion-pion scattering amplitudes are evaluated. Numerical 
results are presented with the choice of $M_{KK}=0.94 GeV$ and $N_c=11$. The results 
are compared with the amplitudes calculated using the experimental phase shifts. The 
agreement with the experimental data is found to be satisfactory. 

\vspace{1.0cm}

{\noindent{\bf{PACS}\ Classification: 11.25.-w; 11.25.Uv}}  
\newpage 
{\noindent{\bf{I.Introduction}}} 

\vspace{0.5cm}

A study of pion-pion scattering is one of the methods to investigate the nature of the 
strong interactions. The standard approach at low energies is the use of the chiral 
perturbation theory, which enables one to calculate the scattering amplitudes near 
threshold [1]. The pion physics at low energies is described by the effective chiral 
lagrangian, 
\begin{eqnarray}
{\cal{L}}_{eff}&=&-\frac{f^2_{\pi}}{4}\ Tr({\partial}_{\mu}U\ {\partial}^{\mu} 
U^{\dagger})+Tr[M(U+U^{\dagger})],
\end{eqnarray}
where $U\ =\ e^{\frac{2i}{f_{\pi}} \Pi(x)}$, $\Pi(x)$ is the pion field, 
$f_{\pi}\simeq 95 MeV$, is the pion decay constant and $M=\frac{f_{\pi}^2m_{\pi}^2}{4}$. 
Weinberg [2] obtained the 
pion-pion scattering amplitude from (1) as 
\begin{eqnarray}
A(s,t,u)&=& \frac{s-m^2_{\pi}}{f^2_{\pi}},
\end{eqnarray}
where $s,t,u$ are the Mandelstam variables satisfying $s+t+u=4m^2_{\pi}$. 

\vspace{0.5cm}

The real scattering amplitudes of definite isospin $R^I(s,t,u)$ are given by 
\begin{eqnarray} 
R^0(s,t,u)&=&3A(s,t,u)+A(t,s,u)+A(u,t,s), \nonumber \\
R^1(s,t,u)&=&A(t,s,u)-A(u,t,s), \nonumber \\
R^2(s,t,u)&=&A(t,s,u)+A(u,t,s),
\end{eqnarray}
where, 
\begin{eqnarray}
s&=&4({\vec{p}}_{\pi}^2+m^2_{\pi}), \nonumber \\
t&=&-2{\vec{p}}_{\pi}^2 (1-cos{\theta}), \nonumber \\
u&=&-2{\vec{p}}_{\pi}^2(1+cos{\theta}). \nonumber 
\end{eqnarray}
${\vec{p}}_{\pi}$ is the pion momentum in the center of mass frame and 
$\theta$ is the scattering angle. The partial wave isospin amplitudes are given by 
\begin{eqnarray}
T^I_{\ell}(s)&=&\frac{1}{64\pi} \sqrt{(1-4\frac{m^2_{\pi}}{s})}\int_{-1}^1 
dcos{\theta}\ P_{\ell}(cos{\theta})\ R^I(s,t,u), 
\end{eqnarray}
with $I=0,1,2$. The current-algebra result (2) gives $T^0_0$ rising almost linearly 
with $\sqrt{s}$ thereby violating the unitarity bound, namely, $|T^I_{\ell}|\leq 
\frac{1}{2}$ even below 1GeV. Dispersion theoretic study by Roy [3] gave the result 
for the above amplitudes consistent with the unitarity bound. However, the various 
parameters involved are to be determined by the experimental data. 

\vspace{0.5cm}

Weinberg [1], in his phenemenological approach, suggested adding two more terms to 
(1) involving four pion field (four derivative contact terms) with arbitrary 
coefficients as 
\begin{eqnarray}
{\cal{L}}_{Weinberg}&=& {\cal{L}}_{eff}+a\ Tr({\partial}_{\mu}U 
{\partial}_{\nu}U^{\dagger}\ {\partial}^{\mu}U {\partial}^{\nu}U^{\dagger}) 
\nonumber \\
&+& 
b\ Tr ({\partial}_{\mu}U {\partial}^{\mu}U^{\dagger}\ {\partial}_{\nu}U  
{\partial}^{\nu}U^{\dagger}),
\end{eqnarray} 
where $a$ and $b$ are dimensionless constants and ${\cal{L}}_{eff}$ is given by (1). The effect of the additional terms in (5) on pion-pion scattering  
has been studied by Sannino and Schechter [4]. They required, arbitrarily, the 
contribution from these terms to be zero at threshold and consequently chose 
$b=-a$ and found for $a\simeq 1\times 10^{-3}$. The four derivative contact terms' 
contribution pull the $T^0_0$ curve to avoid violation of unitarity below  
$\sqrt{s}=1GeV$. While this feature is encouraging, the choice $b=-a$ and the 
numerical value for $a$ are arbitrary. Various phenemenological studies [5,6,7,8]
 have been successful in obtaining the behaviour of the scattering amplitude, 
making use of the experimental data on phase shifts and are consistent with 
unitarity. 

\vspace{0.5cm}

Recently, a holographic dual of QCD with $N_f$ massless quarks, using $D4/D8-
{\bar{D8}}$ brane configuration in Type II-A string theory, with in the 
frame work of AdS/CFT [9,10], has been proposed by Sakai and Sugimoto [11,12] (
hereafter referred to as SS-1 and SS-2 respectively) describing low energy 
phenomena of large $N_c$ QCD. The pion effective action obtained in this model is the 
Skyrme action. The chiral lagrangian derived from SS-1 and SS-2 is found to describe the
axial coupling $g_A$ and the electromagnetic form factors of the nucleon [13,14,15,16]. 
Further, the spontaneous breaking of the chiral symmetry has a geometrical meaning in 
this model. As the model of SS-1, SS-2 describes the low energy QCD in the large 
$N_c$ limit, it is worthwhile to examine this model to understand the low energy 
pion-pion scattering, in particular to derive (5) from this model. 

\vspace{0.5cm}

It is the purpose of this paper to derive the lagrangian (5) from a holographic 
dual of QCD of SS-1 and SS-2. Though pion-pion scattering was considered in SS-2, 
their chiral lagrangian was just (1). By including the $(\alpha')^4$ corrections to 
the non-Abelian DBI action on the $D8$-brane, we obtain (5) with the coefficients $a$ 
and $b$ determined in terms of $g^2_{YM}$ in the large $N_c$ limit. Our results are: 
\begin{eqnarray}
a&=& -\frac{1}{2}C_4, \nonumber \\
b&=&-C_4, \nonumber \\
C_4&=&\frac{1}{g^2_{YM}}\ 0.5865\times 10^{-3}. 
\end{eqnarray}
Clearly, the arbitrariness in choosing $a,b$ in [4] is removed and the coefficients 
are determined in terms of $g^2_{YM}$. Our results show that $b\neq -a$. We have 
evaluated the partial wave scattering amplitudes in (4) using the value of $g^2_{YM}$ 
at large $N_c$ and taking the Kaluza-Klein scale in SS-1 to be that of proton mass. 
Our results respect the unitarity bound for the amplitudes for $N_c\geq 11$ and 
are in good qualitative agreement with the experimental data. 

\newpage 

{\noindent{\bf{II. Brief Review of Sakai-Sugimoto model}}} 

\vspace{0.5cm} 

The model in SS-1 (SS-2) starts with the near horizon limit of a configuration of 
$N_c$ D4-branes wrapping a circle in the 4-direction and the $N_f$ $D8$ and $N_f$ 
${\bar{D8}}$ branes placed at the anti-podal points of this circle. The $D4$ 
background consists of $N_c$ flat D4-branes with one spatial world-volume 
direction compactified on $S^1$ and is given by the supergravity solution of 
Kruczenski, Mateos, Myers and Winters [17]. The space-time extension of the 
D4-brane and the $D8/{\bar{D8}}$-branes can be neatly given by the schematic diagram 
below.

\vspace{0.5cm}

\begin{tabular}{ccccccccccc} 
 &0&1&2&3&4&5&6&7&8&9 \\
 & & & & & & & & & & \\ \hline  
D4&*&*&*&*&*& & & & & \\
$D8/{\bar{D8}}$&*&*&*&*& &*&*&*&*&* 
\end{tabular} 

\vspace{0.5cm}

Here the $*$ denotes the extended directions of the branes. The $x^4$ direction is 
compactified on a circle of radius $M^{-1}_{KK}$ ($M_{KK}$ is the Kaluza-Klein 
mass scale) with anti-periodic boundary condition for the fermions. The effect of this 
is to make the fermions massive, thereby completely breaking 
the supersymmetry. The massless modes of the 4-4 strings are the gauge fields, 
$A_{\mu}^{D4}(\mu =0,1,2,3)$, scalar $A_4^{D4}$ and ${\Phi}^i(i=5,6,7,8,9)$, all in the 
adjoint representation of the gauge group $U(N_c)$. It has been argued in [11,12] that 
the mass terms of the scalar $A_4^{D4}$ and ${\Phi}^i$ are produced by one-loop 
corrections and their trace parts (massless) do not play important role in the low 
energy physics. Thus at energies lower than $M_{KK}$, we have pure Yang-Mills theory 
in $(1+3)$ dimensions. 

\vspace{0.5cm}

Next one introduces $N_f$ flavours of quarks by placing a stack of $N_f$ overlapping 
$D8$ and $\bar{D8}$ branes at the anti-podal points of $S^1$ (compactified $x^4$). 
From the $4-8$ and $4-\bar{8}$ strings (the open strings with one end attached to the $D4$
-brane and the other end to the $D8$-brane and similarly with $\bar{D8}$-brane) we 
obtain $N_f$ flavors of massless fermions in the fundamental representation of the 
$U(N_c)$ gauge group. These fermions are interpreted as quarks in QCD. The chirality 
of the fermion created by the $4-8$ strings is opposite to that created by the 
$4-\bar{8}$ strings (as $\bar{D8}$-brane is $D8$-brane with opposite orientation). 
So the $U(N_f)_{D8}\times U(N_f)_{\bar{D8}}$ gauge symmetry of the $N_f\ D8/{\bar{D8}}$
-pair is interpreted as the $U(N_f)_L\times U(N_f)_R$ chiral symmetry of QCD. 

\vspace{0.5cm}

The compactification of $x^4$ which is responsible for breaking the supersymmetry 
(with antiperiodic boundary condition for fermions) plays a crucial role in 
getting a geometric understanding of chiral symmetry breaking. To see this, we 
consider the $D4$-background supergravity solution [17] with $x^4$ as $\tau$: 
\begin{eqnarray}
ds^2={\Big(\frac{U}{R}\Big)}^{\frac{3}{2}}({\eta}_{\mu\nu}dx^{\mu}dx^{\nu}
+f(U){d\tau}^2)+{\Big(\frac{R}{U}\Big)}^{\frac{3}{2}}(\frac{dU^2}{f(U)}+U^2
d{\Omega}_4), \nonumber \\
e^{\phi}=g_s{\Big(\frac{U}{R}\Big)}^{\frac{3}{4}};F_4=dc_3=\frac{2\pi N_c}{V_4}
{\epsilon}_4;f(U)=1-\frac{U^3_{KK}}{U^3},
\end{eqnarray}
where $x^{\mu}(\mu=0,1,2,3)$ and $\tau$ are the directions along which the 
$D4$-brane is extended, $d{\Omega}_4,{\epsilon}_4, V_4=\frac{8{\pi}^2}{3}$ are 
the line element, volume form and the volume of unit $S^4$, $R(=R_{D4})$ and 
$U_{KK}$ are parameters and the coordinate $U$ is taken to be bounded from below 
($U\geq U_{KK}$). $U=U_{KK}$ corresponds to a horizon in the supergravity 
solution. $g_s(=e^{<\phi>})$ is the string coupling and $R^3=\pi g_s N_c 
{\ell}^3_s$, ${\ell}_s$ being the string length. At $U=U_{KK}$ there is a 
conical singularity in (7). The conical singularity is avoided by using the fact 
that $\tau (=x^4)$ is periodic {\it{with period}} $\delta \tau=2\pi\ \frac{2 
R^{\frac{3}{2}}} {U^{\frac{1}{2}}_{KK}}$.
{\footnote{Let $y=h(U_{KK})\sqrt{f(U)}$. 
Then, $dU=\frac{2\sqrt{f(U)}dy}{h(U_{KK})(\frac{df}{dU})}$. The metric at $U=
U_{KK}$ becomes 
\begin{eqnarray}
(ds)^2|_{U=U_{KK}}&=&{\Big(\frac{U_{KK}}{R}\Big)}^{\frac{3}{2}}[{\eta}_{\mu\nu}
dx^{\mu}dx^{\nu}+\frac{y^2}{h^2(U_{KK})} d{\tau}^2+\frac{4R^3}{U^3_{KK}
h^2(U_{KK})(\frac{df}{dU})^2|_{U=U_{KK}}} dy^2 \nonumber \\
&+&\frac{R^3}{U_{KK}}d{\Omega}^2_4]. \nonumber 
\end{eqnarray}
Let us set $\frac{4R^3}{U^3_{KK}h^2(U_{KK})(\frac{df}{dU})^2|_{U=U_{KK}}}=1$, 
determining $h(U_{KK})$ as $\frac{2R^{\frac{3}{2}}}{3U^{\frac{1}{2}}_{KK}}$.
Then,
\begin{eqnarray}
(ds)^2|_{U=U_{KK}}&=&{\Big(\frac{U_{KK}}{R}\Big)}^{\frac{3}{2}}[{\eta}_{\mu\nu}
dx^{\mu}dx^{\nu}+\frac{y^2}{h^2(U_{KK})}\ d{\tau}^2+dy^2+\frac{R^3}{U_{KK}}
d{\Omega}^2_4]. \nonumber 
\end{eqnarray}
Now, consider $dy^2+\frac{y^2}{h^2(U_{KK})}d{\tau}^2$. Assuming $\tau$ periodic 
with period $2\pi \beta$, let $\phi={\beta}^{-1}\tau$, so that $\phi$ has a period 
$2\pi$. Then, $dy^2+\frac{y^2}{h^2(U_{KK})} d{\tau}^2
\ =\ dy^2+{\alpha}^2y^2 d{\phi}^2$, with $\alpha=\frac{\beta}{h(U_{KK})}$. This 
is a standard cone with singularity at $y=0$. When $\alpha=1$, we have $dy^2+
y^2d{\phi}^2=dX^2+dY^2$, the conical singularity disappears and the period of 
$\tau$ is $2\pi \beta=2\pi h(U_{KK})$ which is $2\pi \frac{2R^{\frac{3}{2}}}{
U^{\frac{1}{2}}_{KK}}$. 
}} 
The Kaluza-Klein mass scale then is $M_{KK}=\frac{2\pi}{\delta \tau}=\frac{
3U^{\frac{1}{2}}_{KK}}{2R^{\frac{3}{2}}}$, below which the dual gauge theory 
is effectively the same as 4-dimensional YM theory, with $g^2_{YM}=4{\pi}^2
g_s{\ell}_s\frac{1}{\delta \tau}$. The parameters $R,U_{KK}$ and $g_s$ are given 
by 
\begin{eqnarray}
R^3\ =\ \frac{g^2_{YM}N_c{\ell}^2_s}{2M_{KK}} &;& U_{KK}\ =\ \frac{2}{9}
g^2_{YM}N_cM_{KK}{\ell}^2_s, \nonumber \\
g_s\ =\ \frac{g^2_{YM}}{2\pi M_{KK}{\ell}_s} &;& M_{KK}\ =\ \frac{3U^{\frac{1}{2}}
_{KK}}{2R^{\frac{3}{2}}}. 
\end{eqnarray} 

\vspace{0.5cm}

One introduces the flavors by placing a stack of $N_f$ overlapping $D8$ and 
$\bar{D8}$-branes at the anti-podal points of the compactified $S^1$, producing 
the global $U(N_f)_L\times U(N_f)_R$ chiral symmetry which is visible on the $D8$ 
and $\bar{D8}$-branes as chiral gauge symmetry. In the probe limit $N_f<<N_c$, 
the back-reaction of the $D8$-branes (and $\bar{D8}$-branes) on the background 
(7) is negligible. The induced metric on the $D8$-brane embedded in the $D4$ 
background (7) with $U=U(\tau)$ is 
\begin{eqnarray}
(ds)^2_{D8}&=&{\Big(\frac{U}{R}\Big)}^{\frac{3}{2}}{\eta}_{\mu\nu}dx^{\mu}
dx^{\nu}+\{ {\Big(\frac{U}{R}\Big)}^{\frac{3}{2}}f(U)+{\Big( \frac{R}{U}\Big)}
^{\frac{3}{2}}\ \frac{ {U'}^2}{f(U)}\} d{\tau}^2 \nonumber \\
&+&{\Big(\frac{R}{U}\Big)}^{\frac{3}{2}} U^2 d{\Omega}^2_4, 
\end{eqnarray}
where $U'=\frac{dU}{d\tau}$. 

\vspace{0.5cm}

The submanifold of the background (7) spanned by $\tau$ and $U$ has the geometry 
of a cigar where the minimum value of $U$ at the tip of the cigar is $U_{KK}$. 
The $D8$ and $\bar{D8}$-branes placed at the antipodals of $S^1$ are well 
separated giving the $U(N_f)_L\times U(N_f)_R$ chiral symmetry. If the $D8$ and 
$\bar{D8}$-branes smoothly join at some point $U=U_0>U_{KK}$, then the chiral 
symmetry $U(N_f)_L\times U(N_f)_R$ spontaneously breaks to $U(N_f)_{L+R}$. This 
can be seen by following the curve $U(\tau)$ to $U=U_0$ at which $U'=0$. To see 
this, we consider the Dirac-Born-Infeld (DBI) action on the $D8$-brane. The 
Chern-Simons term does not affect the solution of the equations of motion 
[11,12,16,18].The DBI-action on the $D8$-brane is 
\begin{eqnarray}
S&=&T_8\int d^4x d\tau d^4\Omega\ e^{-\phi} \ \sqrt{-det g_i},
\end{eqnarray} 
where $T_8=\frac{1}{(2\pi)^8{\ell}^9_s}$ is the tension of the $D8$-brane and 
$g_i$ is the induced metric given by (9). Using (9), we find (10) as 
\begin{eqnarray}
S&=&\frac{ {\tilde{T}}_8}{g_s}\int d\tau \ U^4\ \sqrt{f(U)+{\Big(\frac{R}{U}
\Big)}^3 \frac{ {U'}^2}{f(U)}}, 
\end{eqnarray}
where ${\tilde{T}}_8$ includes the integration of all coordinates except $\tau$. 
As the integrand in (11) is independent of $\tau$, the equation of motion gives 
the 'energy' conservation as 
\begin{eqnarray}
\frac{U^4 f(U)}{\sqrt{f(U)+{\Big(\frac{R}{U}\Big)}^3 \frac{ {U'}^2}{f(U)}}}
&=&constant \ =\ U^4_0\sqrt{f(U_0)},
\end{eqnarray}
where we assume that there is a point $U_0$ at which the profile $U(\tau)$ has 
a minimum, $U'(U=U_0)=0$. The solution in which as $U\rightarrow \infty$, $\tau$ 
goes to $0,L \ (say)$ has been analysed by Aharony, Sonnenschein and Yankielowicz 
[18] and for an asymptotic separation of $L$, the configuration stretches down to 
a minimum at $U=U_0\ >\ U_{KK}$. At $U=U_0$, the $D8$-branes and the $\bar{D8}$-
branes overlap, breaking the chiral symmetry $U(N_f)_L\times U(N_f)_R$ to 
$U(N_f)_{L+R}$. Thus, the chiral symmetry breaking has a geometric description, 
namely, the $D8$-branes and $\bar{D8}$-branes meeting at $U=U_0>U_{KK}$.
Recently, Bergman, Seki and Sonnenschein [26] and Dhar and Nag [26] have modified 
SS-1 model by incorporating the open string tachyon between $D8$ and $\bar{D8}$ 
branes, particularly important when they meet and showed that tachyon condensation 
is responsible for the spontaneous breaking of chiral symmetry. 

\vspace{0.5cm}

In order to describe the $D8$-branes in the probe approximation, it is found to be 
convenient to introduce new coordinates $(y,z)\ =\ (rcos{\theta},rsin{\theta})$ with 
$U_{KK}r^2=U^3-U^3_{KK}$ and $\theta=\frac{3U^{\frac{1}{2}}_{KK}}{2R^{\frac{3}{2}}}
\tau$ so that the metric written in $(y,z)$ is smooth and consider the probe 
$D8$-brane placed at $y=0$ extended along the $z$ direction. This brane configuration 
corresponds to a $D4/D8/\bar{D8}$ system representing $U(N_c)$ QCD with $N_f$ 
massless flavors. (see SS-1 and SS-2 for details). This system has $SO(5)$ symmetry 
corresponding to the rotations of $S^4$. As QCD does not have $SO(5)$ symmetry, 
states invariant under $SO(5)$ rotations alone are considered. The induced metric 
(9) can be written as [11,12] 
\begin{eqnarray}
(ds)^2_{D8}&=&{\Big(\frac{U(z)}{R}\Big)}^{\frac{3}{2}} {\eta}_{\mu\nu} dx^{\mu}
dx^{\nu}+\frac{4}{9}{\Big(\frac{R}{U(z)}\Big)}^{\frac{3}{2}}\ \frac{U_{KK}}
{U(z)}\ dz^2 \nonumber \\
&+&{\Big(\frac{R}{U(z)}\Big)}^{\frac{3}{2}} U^2(z) d{\Omega}^2_4,
\end{eqnarray}
with $U^3(z)=U^3_{KK}+U_{KK}z^2$, and the $D8$-brane extends along $x^{\mu}
(\mu=0,1,2,3)$ and $z$ directions, wrapping around $S^4$. The gauge field on the 
$D8$-brane has nine components, $A_{\mu}(\mu=0,1,2,3),\ A_z,\ A_i(i=5,6,7,8)$ and 
to focus on the $SO(5)$ singlet states, $A_i'$s are set equal to zero. Further, 
$A_{\mu}$ and $A_z$ are taken to be independent of the coordinates on $S^4$. An 
effective action for mesons is obtained in SS-1 (SS-2) from the non-Abelian DBI-action
 of the probe $D8$-brane in a gauge $A_z=0$. The gauge field $A_{\mu}(x^{\mu},z)$ is 
 expanded as 
\begin{eqnarray}
A_{\mu}(x^{\mu},z)&=&U^{-1}(x){\partial}_{\mu}U(x)\ {\psi}_+(z)+\sum_{n\geq 1}
B^{(n)}_{\mu}(x){\psi}_n(z),
\end{eqnarray}
where 
\begin{eqnarray}
{\psi}_+(z)&=&\frac{1}{2}+\frac{1}{\pi}tan^{-1}(\frac{z}{U_{KK}}), \nonumber \\
U(x)&=&exp(\frac{2i}{f_{\pi}} \Pi(x),
\end{eqnarray}
with $\Pi(x)$ as the pion field. The mode function ${\psi}_n(z)$ satisfies
\begin{eqnarray}
-U(z){\partial}_z\{\frac{U^3(z)}{U^2_{KK}} {\partial}_z{\psi}_n\}&=&
{\lambda}_n {\psi}_n, 
\end{eqnarray}
with the normalization 
\begin{eqnarray}
{\tilde{T}}_8 (2\pi {\alpha'})^2 R^3\int dz \frac{1}{U(z)} \ {\psi}_n(z)
{\psi}_m(z) &=& {\delta}_{nm},
\end{eqnarray}
where 
\begin{eqnarray}
{\tilde{T}}_8&=&\frac{2}{3}R^{\frac{3}{2}} U^{\frac{1}{2}}_{KK} T_8 V_4 g^{-1}_s. 
\end{eqnarray}

\vspace{0.5cm}

An effective 4-dimensional action for the pions was obtained by SS-1 (SS-2) from the 
DBI action on the $D8$-brane, by omitting the second term in (14) and integrating 
the $z$-coordinate as 
\begin{eqnarray}
S&=&\int d^4x \{ \frac{f^2_{\pi}}{4} Tr (U^{-1}(x) {\partial}_{\mu}U(x))^2
\nonumber \\
&+&\frac{1}{32e^2} Tr [U^{-1}(x){\partial}_{\mu}U(x), U^{-1}(x){\partial}_{\nu}
U(x)]^2\}, 
\end{eqnarray}
where one sets 
\begin{eqnarray}
\frac{1}{54{\pi}^4}(g^2_{YM}N_c) M^2_{KK} N_c &=& f^2_{\pi}, \nonumber \\
\frac{27{\pi}^7}{2b}\ \frac{1}{(g^2_{YM}N_c)N_c}&=& e^2; \ \ \ b=15.24. 
\end{eqnarray}
This action coincides with the Skyrme action [19], upon identifying $f_{\pi}$ as the 
pion decay constant. The action for vector mesons along with their properties were 
obtained in [11,12]. 

\vspace{0.5cm}

In [11], pion-pion scattering, pion-vector meson interaction 
and decays of vector mesons were studied. In the study of pion-pion scattering, it 
was found in [12] that when the infinite tower of massive vector mesons were included, 
scattering was governed only by the chiral lagrangian (the first term in (19)) for 
pions. This term is exactly the same as (1) and so the 'Weinberg terms' in (5) are 
not obtained in the above holographic description. Phenemenologically, the pion-
pion scattering has been described by including scalar and $\rho$-meson exchanges, 
as in [4] and in [20]. Recently, Harada, Matsuzaki and Yamawaki [21] studied the 
implications of the Sakai-Sugimoto model with hidden local symmetry [22]. By 
keeping $n=1$ vector meson (rho meson) in (14) and integrating the remaining 
($n>1$) mesons, they obtain a lagrangian containing pions, scalars and rho meson 
fields. This gives a better estimate of $\rho-\pi-\pi$ coupling. By using this 
lagrangian to describe pion-pion scattering, one can relate the phenemenological 
approach [4,20] to SS-1/2 model. In [21], (their eqn.27) the resulting lagrangian 
contains besides the Skyrme term, $O(p^4)$ terms. They [21] consider quantum loop 
corrections to describe low energy pion-pion scattering. We are considering the tree 
level of the SS-1 (SS-2) but include the ${\alpha'}^4$ terms to describe the low energy 
pion-pion scattering. This in effect captures the results of the quantum calculations of 
[21]. This motivates us to consider 
${\alpha'}^4$ terms in the DBI action on the D8-branes. 
In the next section, we show that 
the 'Weinberg terms' in (5) are obtained from the DBI-action on the $D8$-brane by 
considering $(\alpha')^4$ terms. Further, the coefficients $a$ and $b$ will be 
determined in terms of $g^2_{YM}$. 

\vspace{1.0cm}

{\noindent{\bf{III. DBI-Action on $D8$-Brane}}} 

\vspace{0.5cm}

The non-Abelian DBI-action on the $D8$-brane is 
\begin{eqnarray}
S^{DBI}_{D8}&=&T_8\int d^9x\ e^{-\phi}\ Str\sqrt{-det(g_{MN}+(2\pi\alpha')F_{MN})},
\end{eqnarray}
where $\alpha'={\ell}^2_s$ is the Regge slope parameter, $T_8=\frac{1}{(2\pi)^8
{\ell}^9_s}$ is the tension of the $D8$-brane. In (21), $g_{MN}$ is the induced 
metric given in (13) with $M,N$ taking values $0,1,2,3,\cdots 8$, $Str$ is the 
symmetric trace and 
\begin{eqnarray}
F_{MN}&=&{\partial}_MA_N-{\partial}_NA_M+i[A_M,A_N], 
\end{eqnarray}
where $A_M$ is the gauge field. The gravitational energy of the $D8$-brane is 
$S^{DBI}_{D8}|_{A_M=0}$ and subtracting this as vacuum energy relative to the 
gauge sector, we have 
\begin{eqnarray}
{\tilde{S}}^{DBI}_{D8}=S^{DBI}_{D8}-S^{DBI}_{D8}|_{A_M=0} &=&T_8\int d^9x e^{-\phi} 
Str\{\sqrt{-det(g_{MN}+(2\pi\alpha')F_{MN})} \nonumber \\ 
&-&\sqrt{-det g_{MN}}\}. 
\end{eqnarray} 

\vspace{0.5cm} 

The right hand side above is expanded as in [23] to give 
\begin{eqnarray}
{\tilde{S}}^{DBI}_{D8}&=&\frac{T_8}{4}(2\pi\alpha')^2\int d^9x e^{-\phi}\ \sqrt{
-det g_{MN}} Tr\Big( F_{MN}F^{MN} \nonumber \\
&-&\frac{1}{3}(2\pi\alpha')^2\{F_{MN}F^{RN}F^{ML}F_{RL}+\frac{1}{2}F_{MN}F^{RN}F_{RL}
F^{ML} \nonumber \\
&-&\frac{1}{4}F_{MN}F^{MN}F_{RL}F^{RL}-\frac{1}{8}F_{MN}F^{RL}F^{MN}F_{RL}
\} + O({\alpha'}^4)\Big). 
\end{eqnarray}
In above, the ${\alpha'}^4$ terms are contained in $\{ \}$. Such terms have been 
earlier 
considered in the fluctuation analysis in the background intersecting branes [24] 
and in the computation of soliton mass [15] in the Sakai-Sugimoto model. We [25] 
have included them in our earlier study of pion-pion scattering. 
 
\vspace{0.5cm}

Now, focussing on $SO(5)$ singlets by taking $A_i=0$ and $A_{\mu}$ independent of 
the coordinates of $S^4$, the coordinates of $S^4$ are 
integrated to give 
\begin{eqnarray}
{\tilde{S}}^{DBI}_{D8}&=&\Big( \frac{2}{3}T_8R^{\frac{3}{2}}U^{\frac{1}{2}}_{KK}
V_4g^{-1}_s\Big) [(2\pi \alpha')^2\int d^4x dz Tr\{\frac{1}{4}\frac{R^3}{U(z)}
{\eta}^{\mu\nu}{\eta}^{\lambda\sigma}F_{\mu\lambda}F_{\nu\sigma}  \nonumber \\ 
&+&\frac{9}{8}\frac{U^3(z)}{U_{KK}}{\eta}^{\mu\nu}F_{\mu z}F_{\nu z}\} \nonumber \\
&-&\frac{1}{12}(2\pi \alpha')^4\ \int d^4x dz Tr\{\frac{R^6}{U^4(z)} {\eta}^{\mu\nu}
{\eta}^{\lambda\sigma}{\eta}^{\rho\delta}{\eta}^{\alpha\beta}(F_{\mu\lambda}
F_{\delta\sigma}F_{\nu\beta}F_{\rho\sigma} \nonumber \\
&-&\frac{1}{8}F_{\mu\lambda}F_{\beta\delta}F_{\nu\sigma}F_{\alpha\rho}+\frac{1}{2}
F_{\mu\lambda}F_{\beta\sigma}F_{\alpha\rho}F_{\nu\delta}-\frac{1}{4}
F_{\mu\lambda}F_{\nu\sigma}F_{\alpha\rho}F_{\beta\delta}) \nonumber \\
&+&\frac{9}{4}\frac{R^3}{U_{KK}}{\eta}^{\mu\nu}{\eta}^{\lambda\sigma}
{\eta}^{\alpha\beta}(2F_{\mu\lambda}F_{\beta\sigma}F_{\nu z}F_{\alpha z}+
F_{\mu\lambda}F_{\sigma z}F_{\nu\beta}F_{\alpha z} \nonumber \\
&+&F_{\mu z}F_{\sigma \nu}F_{\beta z}F_{\lambda z}-\frac{1}{2}F_{\mu\lambda}
F_{\beta z}F_{\nu\sigma}F_{\alpha z}+\frac{1}{2}F_{\mu\lambda}F_{\beta\sigma}
F_{\alpha z}F_{\nu z} \nonumber \\
&+&\frac{1}{2} F_{\mu\lambda}F_{\sigma z}F_{\alpha z}F_{\nu\beta}+\frac{1}{2}
F_{\mu z}F_{\beta z}F_{\alpha\lambda}F_{\nu\sigma}+\frac{1}{2}F_{\beta z}
F_{\mu z}F_{\nu\sigma}F_{\alpha\lambda} \nonumber \\
&-&F_{\mu\lambda}F_{\nu\sigma}F_{\alpha z}F_{\beta z}) \nonumber \\
&+&\frac{81 U^4(z)}{16 U^2_{KK}}{\eta}^{\mu\nu}{\eta}^{\lambda\sigma}(\frac{1}{2}
F_{\mu z}F_{\sigma z}F_{\nu z}F_{\lambda z}+F_{\mu z}F_{\sigma z}
F_{\lambda z}F_{\nu z})\}], 
\end{eqnarray} 
where the prefactor is just ${\tilde{T}}_8$ given in (18) and we have displayed the 
$(\alpha')^4$-correction terms in the second $\{ \}$ expression. 

\vspace{0.5cm}

The above action (25) is general upto $(\alpha')^4$ terms and the gauge fields  
$A_{\mu}(x,z)$ and $A_z(x,z)$ can be expanded using complete sets of functions of $z$.
A 4-dimensional effective action is obtained by integrating $z$. We now work in 
$A_z=0$ gauge. We are interested in the terms in (25) contributing to pion-pion 
scattering only, in particular of the form (5). 
In (25) the first two terms have been considered in SS-2 [12] for 
low energy $\pi-\pi$ scattering. The use of (14) led to the result that the low 
energy behaviour of the $\pi-\pi$ scattering amplitude is goverened only by the 
${\Pi}^4$ vertex coming from the lowest derivative term in $U$, namely $(U^{-1}
{\partial}_{\mu}U)^2$, giving the first term in (5).  
Now, when (14) is substituted in the $(\alpha')^4$ terms in 
(25), it is seen that the last two terms of (5) come from the last two terms of 
(25) without involving $B_{\mu}^n(x)$. So, as far as a description of pion-pion 
scattering is concerned, the above is equivalent to using (14) without the 
second term and consistently omitting the Skyrme term in the final expression. 
In view of this we consider $A_{\mu}(x,z)$ as 
\begin{eqnarray}
A_{\mu}(x,z)&=&U^{-1}(x){\partial}_{\mu}U(x)\ {\psi}_+(z),
\end{eqnarray}
where ${\psi}_+(z)$ and $U(x)$ are 
given in (15). From (26), it follows 
\begin{eqnarray}
F_{\mu\nu}(x,z)&=&[U^{-1}(x){\partial}_{\mu}U(x),U^{-1}(x){\partial}_{\nu}U(x)]
{\psi}_+(z)\ ({\psi}_+(z)-1), \nonumber \\
F_{z\mu}(x,z)&=&U^{-1}(x){\partial}_{\mu}U(x)\ {\partial}_z{\psi}_+(z) \equiv 
U^{-1}(x){\partial}_{\mu}U(x) {\hat{\phi}}_0(z),
\end{eqnarray}
where 
\begin{eqnarray}
{\hat{\phi}}_0(z)&=&\frac{U^2_{KK}}{\pi}\ \frac{1}{U^3(z)},
\end{eqnarray}
which follows from (15) and the relation $U^3(z)=U^3_{KK}+U_{KK}z^2$ below (13). 
Substituting (27) and (28) in (25), we encounter the 
following $z$-integrals which are evaluated as 
\begin{eqnarray}
\int \frac{ {\psi}^2_+(z)({\psi}_+(z)-1)^2}{U(z)}dz &=&\frac{15.246}{ {\pi}^4}, 
\nonumber \\
\int U^3(z) {\hat{\phi}}^2_0(z)dz &=&\frac{U^2_{KK}}{\pi}, \nonumber \\
\int U^4(z) {\hat{\phi}}^4_0(z) dz&=&\frac{1.275\ U_{KK}}{ {\pi}^4}, \nonumber \\
\int {\psi}^2_+(z)({\psi}_+(z)-1)^2\ {\hat{\phi}}^2_0(z) dz&=& \frac{7.455}{
U_{KK}{\pi}^6}, \nonumber \\
\int \frac{ {\psi}^4_+(z)({\psi}_+(z)-1)^4}{U^4(z)} dz&=& \frac{43.738}{U^3_{KK}
{\pi}^8}. 
\end{eqnarray} 

\vspace{0.5cm}

Introducing ${\ell}_{\mu}\ =\ U^{-1}(x){\partial}_{\mu}U(x)$, the action (25) after the 
$z$ integration using (29) becomes
\begin{eqnarray}
{\tilde{S}}^{DBI}_{D8}&=&\int d^4x\ Tr\{ \frac{f^2_{\pi}}{4} {\ell}_{\mu}{\ell}^{\mu}+
\frac{1}{32e^2}[{\ell}_{\mu},{\ell}_{\nu}]^2\} \nonumber \\
& & \nonumber \\ 
&-&\frac{1}{24}{\tilde{T}}_8(2\pi \alpha')^4\int d^4x\ Tr[{\Big(
\frac{2\times 43.738\ R^6}{ {\pi}^8\ U^3_{KK}}\Big)} {\eta}^{\mu\nu}{\eta}^{\lambda\sigma} 
{\eta}^{\rho\delta}{\eta}^{\alpha\beta} \nonumber \\
& &\{[{\ell}_{\mu},{\ell}_{\lambda}][{\ell}_{\delta},{\ell}_{\sigma}][{\ell}_{\nu},
{\ell}_{\beta}][{\ell}_{\rho},{\ell}_{\alpha}] 
-\frac{1}{8}[{\ell}_{\mu},{\ell}_{\lambda}][{\ell}_{\beta},{\ell}_{\delta}]
[{\ell}_{\nu},{\ell}_{\sigma}][{\ell}_{\alpha},{\ell}_{\rho}] \nonumber \\
&+&\frac{1}{2}[{\ell}_{\mu},{\ell}_{\lambda}][{\ell}_{\beta},{\ell}_{\sigma}] 
[{\ell}_{\alpha},{\ell}_{\rho}][{\ell}_{\nu},{\ell}_{\delta}] 
-\frac{1}{4}[{\ell}_{\mu},{\ell}_{\lambda}][{\ell}_{\nu},{\ell}_{\sigma}] 
[{\ell}_{\alpha},{\ell}_{\rho}][{\ell}_{\beta},{\ell}_{\delta}]\} \nonumber \\
& & \nonumber \\ 
&+&\Big(\frac{9\times 7.455\ R^3}{U^2_{KK}\ {\pi}^6}\Big){\eta}^{\mu\nu}{\eta}^{\lambda
\sigma}{\eta}^{\alpha\beta}\{ 2[{\ell}_{\mu},{\ell}_{\lambda}][{\ell}_{\beta},
{\ell}_{\sigma}]\ {\ell}_{\nu}{\ell}_{\alpha} \nonumber \\ 
&+&[{\ell}_{\mu},{\ell}_{\lambda}]\ {\ell}_{\sigma}[{\ell}_{\nu},{\ell}_{\beta}]\ 
{\ell}_{\alpha}+{\ell}_{\mu}\ [{\ell}_{\sigma},{\ell}_{\nu}]\ {\ell}_{\beta}
[{\ell}_{\lambda},{\ell}_{\alpha}]  
-\frac{1}{2}[{\ell}_{\mu},{\ell}_{\lambda}]\ {\ell}_{\beta}[{\ell}_{\nu},
{\ell}_{\sigma}]\ {\ell}_{\alpha} \nonumber \\
&+&\frac{1}{2}[{\ell}_{\mu},{\ell}_{\lambda}]
[{\ell}_{\beta},{\ell}_{\sigma}]\ {\ell}_{\alpha}{\ell}_{\nu} 
+\frac{1}{2}[{\ell}_{\mu},{\ell}_{\lambda}]\ {\ell}_{\sigma}{\ell}_{\alpha}\ 
[{\ell}_{\nu},{\ell}_{\beta}]+\frac{1}{2}{\ell}_{\mu}{\ell}_{\beta}\ [{\ell}_{\alpha},
{\ell}_{\lambda}][{\ell}_{\nu},{\ell}_{\sigma}] \nonumber \\ 
&+&\frac{1}{2}{\ell}_{\beta}{\ell}_{\mu}\ [{\ell}_{\nu},{\ell}_{\sigma}][{\ell}_{\alpha},
{\ell}_{\lambda}]-[{\ell}_{\mu},{\ell}_{\lambda}][{\ell}_{\nu},{\ell}_{\sigma}]
\ {\ell}_{\alpha}{\ell}_{\beta}\} \nonumber \\
& & \nonumber \\
&+&\Big(\frac{81\times 1.275}{8\ U_{KK} {\pi}^4}\Big){\eta}^{\mu\nu}{\eta}^{\lambda
\sigma}\{\frac{1}{2}{\ell}_{\mu}{\ell}_{\sigma}{\ell}_{\nu}{\ell}_{\lambda}+
{\ell}_{\mu}{\ell}_{\nu}{\ell}_{\sigma}{\ell}_{\lambda}\}], 
\end{eqnarray} 
where we have identified 
\begin{eqnarray}
\frac{9}{2}{\tilde{T}}_8\ (2\pi \alpha')^2\ \frac{U_{KK}}{\pi}&=&f^2_{\pi}, \nonumber \\
\frac{15.246\ R^3}{4{\pi}^4}\ {\tilde{T}}_8\ (2\pi \alpha')^2&=&\frac{1}{32\ e^2},
\end{eqnarray}
which are same as in (20). 

\vspace{0.5cm}

The first two terms in (30) reproduce the Skyrme model with the identification (31).
From (20) (and 
(31)), it is encouraging to observe that the $N_c$-dependence of $f_{\pi}$ and $e$, in 
the large $N_c$ limit for fixed 'tHooft coupling $g^2_{YM}N_c$ goes as $O(\sqrt{N_c})$ 
and $O(\frac{1}{\sqrt{N_c}})$ respectively, in agreement with the results of large $N_c$ 
QCD. The effective action for pions, from the holographic QCD upto $(\alpha')^4$ terms 
is (30). 

\vspace{0.5cm}

Now, using ${\ell}_{\mu}=U^{-1}(x){\partial}_{\mu}U(x)$, the first two terms in (30) 
give (1). The last two terms in (30) give precisely the 'Weinberg terms' in (5) with 
the coefficients $a$ and $b$ {\it{determined}} as 
\begin{eqnarray}
a&=&-\frac{1}{54}\ \Big( \frac{81\times 1.275}{8\ U_{KK}\ {\pi}^4}\Big)\ {\tilde{T}}_8 
(2\pi \alpha')^4, \nonumber \\
& & \nonumber \\
b&=&-\frac{1}{24}\ \Big(\frac{81\times 1.275}{8\ U_{KK}\ {\pi}^4}\Big)\ {\tilde{T}}_8
(2\pi \alpha')^4.
\end{eqnarray}
{\it{This is our main result, namely, the 'Weinberg terms' are obtained from the 
$D8$-brane DBI-action in the Sakai-Sugimoto model with their coefficients determined.}}
Using (8) and the relation 
\begin{eqnarray}
{\tilde{T}}_8(2\pi \alpha')^2=\Big(\frac{2}{3} T_8 R^{\frac{3}{2}} U^{\frac{1}{2}}
_{KK} V_4 g^{-1}_s\Big) (2\pi \alpha')^2 &=& \frac{1}{54{\pi}^3}\ M_{KK} N_c {\ell}^{-2}
_s,
\end{eqnarray}
we find 
\begin{eqnarray}
a&=&-\frac{1.173\times 10^{-3}}{4\ g^2_{YM}}, \nonumber \\
b&=&-\frac{1.173\times 10^{-3}}{2g^2_{YM}},
\end{eqnarray}
which have been displayed earlier in (6). As desired, these coefficients are dimensionless.
The remaining terms in (30) give rise to pion-pion scattering leading to four and six 
pions in the final state. 

\vspace{1.0cm}

{\noindent{\bf{IV. Effective Lagrangian for pions - Low energy pion-pion scattering}}}

\vspace{0.5cm}

A lagrangian describing pions upto four derivatives of $U(x)$ is given by the first 
two terms and last two terms in (30). The second term in (30) is the familiar Skyrme 
term 
introduced by Skyrme to stabilize the soliton. The holographic model reproduces this 
term. This term can also contribute to $\pi-\pi\ \rightarrow \pi-\pi$ scattering. As
explained earlier (paragraph before (26)), in 
[12], it was shown that when massive vector mesons are invoked, the contribution from 
the Skyrme term gets cancelled by the vector meson exchange diagrams. The resulting 
contribution to low energy pion-pion scattering amplitude comes from the four 
derivatives of $U(x)$ in the first and last two terms in (30). These are exactly in 
(5) with the coefficients in (6). There is an attempt to realize pion mass term 
in SS-1 and SS-2 model by introducing instantons in $S^4$ [16]. This, in view of 
Gell-Mann-Oakes-Renner relation suggests quark mass and it is not clear, at present, 
how to introduce quark masses in Sakai-Sugimoto model, though there are attempts in 
this direction [26]. In view of this, we add to the lagrangian for pions 
obtained from SS-1 and SS-2 model, a mass term for pions by hand as in (1).

\vspace{0.5cm}

The pion-pion scattering amplitude from (5) and (34,6) is given by 
\begin{eqnarray}
A(s,t,u)&=&\frac{s-m^2_{\pi}}{f^2_{\pi}}-\frac{2C_4}{f^4_{\pi}}\{(t-2m^2_{\pi})^2 
+(u-2m^2_{\pi})^2+(s-2m^2_{\pi})^2\},
\end{eqnarray}
which can be verified using the results in [4] with $a=b/2$. The partial wave scattering 
amplitudes can be calculated from (3) and (4). It is found that the $I=1$ amplitude 
$R^1(s,t,u)$ is independent of the $(\alpha')^4$ corrections and so we do not consider it 
here. The other amplitudes are found as 
\begin{eqnarray}
T^{I=0}_{\ell=0}(s)&=&\frac{1}{64\pi}\sqrt{1-\frac{4m^2_{\pi}}{s}}[\frac{2}{f^2_{\pi}}
(2s-m^2_{\pi}) \nonumber \\
&-&\frac{10\ C_4}{f^4_{\pi}}\{2(s-2m^2_{\pi})^2+s^2+\frac{1}{3}(s-4m^2_{\pi})^2\}],
\nonumber \\
& & \nonumber \\
T^2_0(s)&=&-\frac{1}{64\pi}\sqrt{1-\frac{4m^2_{\pi}}{s}}[\frac{2(s-2m^2_{\pi})}{
f^2_{\pi}} \nonumber \\
&+&\frac{4\ C_4}{f^4_{\pi}}\{s^2+2(s-2m^2_{\pi})^2+\frac{1}{3}(s-4m^2_{\pi})^2\}], 
\nonumber \\
& & \nonumber \\
T^2_2(s)&=&-\frac{1}{64\pi}\sqrt{1-\frac{4m^2_{\pi}}{s}}\ \frac{8\ C_4}{15\ f^4_{\pi}}
\ (s-4m^2_{\pi})^2, \nonumber \\
& & \nonumber \\
T^0_2(s)&=&-\frac{1}{48\pi}\sqrt{1-\frac{4m^2_{\pi}}{s}}\ \frac{C_4}{f^4_{\pi}}\ 
(s-4m^2_{\pi})^2. 
\end{eqnarray} 

\vspace{0.5cm}

It is seen that the partial wave $\ell=2$ amplitudes are dependent only on the 
${\alpha'}^4$ terms. The dimensionless coupling $C_4$ in (6) is independent of 
$N_c,M_{KK},{\ell}_s$, the parameters of the Sakai-Sugimoto model. It depends only 
on the Yang-Mills coupling $g^2_{YM}$. However, as we are working in the large $N_c$ 
holographic QCD, its numerical value cannot be fixed completely. The scale at which 
the 5-dimensional action is reduced to 4-dimensional action is the Kaluza-Klein 
scale, i.e., $M_{KK}$. In SS-1 and SS-2, it was found that the choice $M_{KK}\simeq 
0.94 GeV$ is appropriate. We use the relation (20) or (31), to express $g^2_{YM}$ as 
\begin{eqnarray}
g^2_{YM}&=&\frac{54 {\pi}^4 f^2_{\pi}}{M^2_{KK} N^2_c}\ =\ \frac{51.43}{N^2_c},
\end{eqnarray}
where $f_{\pi}=93 MeV$ and $M_{KK}=0.94 GeV$ are used. From (37), the values for 
$g^2_{YM}$ at $M_{KK}=0.94 GeV$ depend on $N_c$. The contribution of the ${\alpha'}^4$ 
terms to the low energy pion-pion scattering involves $f_{\pi}^2$ (as given by (30),
(31)) and for its numerical value we have chosen $f_{\pi}\simeq 95 MeV$. The 
contribution of the ${\alpha'}^4$ terms to the low energy pion-pion scattering involves 
$g_{YM}^2$ as given in (34). The Yang-Mills coupling at large $N_c$ and at $M_{KK}=
0.94 GeV$ is given in (37) with in the SS-model and this involves $N_c$. Eqn.37 is 
consistent with the relation $\lambda N_c\simeq 50$ as found in the Table.1 of [14]. 
Instead of the usual practice of large $N_c$ calculations of taking the leading order 
$N_c$ terms in the expansion and setting $N_c=3$ at the end, we have chosen to use 
directly 'large' $N_c$ values, since (34) involves only $g_{YM}^2$ which is 
evaluated at $M_{KK}=0.94 GeV$ in (37). The 'large' value is chosen to be consistent 
with the unitarity bound of the scattering amplitude. By numerical calculations using 
various values of $N_c$ starting from 3, we find the lowest value of $N_c$ which makes 
the scattering amplitude $\leq 0.5$, as 11. 

\vspace{1.0cm}

{\noindent{\bf{V. Numerical Results and Discussion}}}

\vspace{0.5cm}

We have evaluated the partial wave scattering amplitudes in (36) for $\pi-\pi
\rightarrow \pi-\pi$ scattering below $1 GeV$ for various values of $N_c$ and find that 
the lowest $N_c$ value for which the amplitude $T^0_0$ is within the unitarity bound,
i.e., $|T^0_0|\leq 0.5$ is $N_c=11$. With this value, we find $C_4=1.38\times 10^{-3}$
 and $a$ and $b$ are obtained using (6). 

 \vspace{0.5cm}

In Fig.1, we have plotted the amplitudes $T^0_0$ 
with $\sqrt{s}$ evaluated using (1) and (5), using our result for $C_4$ for $N_c=11$. The points $+$ are obtained using (36) with $N_c=3$. They follow nearly 
the current algebra result. They, besides violating the unitarity bound, do not 
agree with the experimental data in Fig.2. 
The 'almost linearly rising' behaviour of the current-algebra result is ameliorated by 
the ${\alpha'}^4$ corrections added as in (36) with $N_c=11$. 

\vspace{0.5cm}

The experimental data on the phase 
shifts ${\delta}^0_0$ are taken from the analysis of Kami\'{n}ski, Pel\'{a}ez 
and Yndur\'{a}in [27].
Using these phase shifts, we have evaluated the amplitude $T^0_0$ and compared with 
our theoretical results with $N_c=11$ in Fig.2. 

\vspace{0.5cm} 

From this figure (Fig.2), it is seen that the experimental 
curve is satisfactorily explained by the theoretical amplitude in (36) with $N_c=3$. 
It is useful 
to compare our theoretical curve with [4] (their Fig.7; with $b=-a$ and for $a=0.5,
0.7,1$ in units of $10^{-3}$). The curves in [4] show the vanishing of the $T^0_0$ 
amplitudes after $\sqrt{s}=1 GeV$ while the experimental amplitudes vanish around 
$\sqrt{s}$ less than 1 GeV in agreement with our theory, bringing in the holographic 
QCD description closer to realistic QCD. 
In [28], 
meson-meson scattering amplitude was considered using non-linear and linear chiral 
lagrangians and a best fit for $T^0_0$ with various values of their $m_{bare}(\sigma)$ 
is obtained. This is fit is in agreement with our theoretical curve in holographic 
QCD approach. 

\vspace{0.5cm}

Now, we consider $\ell=0; I=2$ amplitude $T^2_0$. This evaluated using (36) and plotted 
in Fig.3 along with the results obtained using the experimental phase shifts from [24].
It is seen from the Fig.3, that our theoretical amplitude upto $\sqrt{s}=0.6 GeV$ are 
in reasonable agreement with the experimental amplitude calculated from the phase 
shifts. In this case, the lowest inelastic process is $\pi\pi\rightarrow \pi\pi\rho$ and
 as the $\rho$-meson mass is 770 MeV, the above description is inadequate after this 
 value for $\sqrt{s}$ as we have not taken into account the rho production. 

\vspace{0.5cm}

We now consider $\ell=2\ ;\ I=2$ amplitude $T^2_2$. This is evaluated using (36) with 
$M_{KK}=0.94 GeV$ and $N_c=11$ and is plotted in Fig.4, along with the experimental 
amplitude calculated using the phase shifts from [24]. The agreement with the experiment  is reasonable. The amplitude $T^0_2$ can be calculated from (36). However, as seen 
in [24] (their Fig.7.1c), the phase shifts ${\delta}^0_2$ are near zero till about 
1 GeV. The theoretical values are consistent with zero.  

\vspace{0.5cm}

To summarize, we have considered the holographic large $N_c$ QCD proposed by 
Sakai and Sugimoto and restricting to the pion sector, evaluated the effective 
action upto ${\alpha'}^4$ terms. The effective chiral lagrangian for pions is 
derived and this coincides with the lagrangian suggested by Weinberg. The 
couplings of the four derivative contact terms are determined in terms of $g^2_{YM}$.
In the spirit of large $N_c$, these couplings are re-expressed in terms of the 
pion decay constant, Kaluza-Klein scale and $N_c$. Using $M_{KK}=0.94 GeV$ and 
$f_{\pi}=95 MeV$, we find that $N_c=11$ gives the $T^0_0$ amplitude within the 
unitarity bound. For these values, the partial wave amplitudes are evaluated and 
compared with the results obtained from experimental phase shifts. The agreement 
with the experimental data is found to be satisfactory within about 20 percent. 
It will be interesting to consider baryons within this approach and examine the 
baryon-meson coupling. This work is under progress. 

\vspace{0.5cm}
 
{\noindent{\bf{Acknowledgement}}} 

\vspace{0.5cm}

One of us (K.S.V) acknowledges the warm hospitality at the Chennai Mathematical 
Institute. This research is supported by an operating grant from Natural Sciences 
and Engineering Council of Canada. Useful correspondence with J.R.Pel\'{a}ez is 
acknowledged with thanks. 

\vspace{1.0cm}

{\noindent{\bf{References}}} 

\vspace{0.5cm}

\begin{enumerate}
\item S.Weinberg, Physica {\bf{A96}}, 327 (1979). 
\item S.Weinberg, Phys.Rev.Lett. {\bf{17}}, 616 (1966).
\item S.M.Roy, Phys.Lett. {\bf{36}}, 353 (1971). 
\item F.Sannino and J.Schechter, Phys.Rev. {\bf{D52}}, 96 (1995).
\item B.Ananthanarayanan and P.Buttiker, ``Pion-Pion scattering in Chiral Perturbation 
and Dispersion Relation Theories'', arXiv:hep-ph/9902461.
\item J.Gasser and H.Leutwyler, Ann.Phys. (N.Y). {\bf{158}}, 142 (1984); Nucl.Phys.
{\bf{B250}}, 465 (1985).
\item G.Colangelo, J.Gasser and H.Leutwyler, Nucl.Phys. {\bf{B603}}, 125 (2001).
\item J.R.Pel\'{a}ez and F.J.Yndur\'{a}in, Phys.Rev. {\bf{D71}}, 074016 (2005).
\item J.M.Maldacena, Adv.Theor.Math.Phys. {\bf{2}}, 331 (1998); arXiv:hep-th/9711200.
\item S.S.Gubser, I.R.Klebanov and A.M.Polyakov, Phys.Lett. {\bf{B428}}, 105 (1998);
arXiv:hep-th/9802109. \
      E.Wittn, Adv.Theor.Math.Phys. {\bf{2}}, 253 (1998); arXiv:hep-th/9802150. \
      O.Aharony, S.S.Gubser, J.M.Maldacena, H.Ooguri and Y.Oz, Phys.Rep. {\bf{323}},
       183 (200); arXiv:hep-th/9905111. 
\item T.Sakai and S.Sugimoto, Prog.Theor.Phys. {\bf{113}}, 843 (2005); arXiv:hep-th/
      0412141. 
\item T.Sakai and S.Sugimoto, Prog.Theor.Phys. {\bf{114}}, 1083 (2006); arXiv:hep-th/ 
      0507073.
\item D.K.Hong, M.Rho, H-U.Yee and P.Yi, ``Chiral Dynamics of Baryons from String
Theory'', arXiv:hep-th/0701276. 
\item D.K.Hong, M.Rho, H-U.Yee and P.Yi, ``Dynamics of Baryons from String Theory and 
Vector Dominance'', arXiv: hep-th/07052632.
\item H.Hata, T.Sakai, S.Sugimoto and S.Yamato, ``Baryons from Instantons in holographic  QCD'', arXiv: hep-th/0701280. 
\item K.Nawa, H.Suganuma and T.Kojo, ``Baryons in holographic QCD'', arXiv: hep-th/
0612187. \\
      K.Hashimoto, T.Hirayama and A.Miwa, ``Holographic QCD and Pion Mass'', 
      arXiv: hep-th/0703024. 
\item M.Kruczenski, D.Mateos, R.C.Myers and D.J.Winters, JHEP, {\bf{5}}, 041 (2004);
arXiv: hep-th/0311270.
\item O.Aharony, J.Sonnenschein and S.Yankielowicz, ``A Holographic Model of 
Deconfinement and Chiral Symmetry Restoration'' , arXiv: hep-th/0604161.
\item T.H.R.Skyrme, Proc.Roy.Soc.Lond. {\bf{A260}}, 127 (1961); Nucl.Phys.
{\bf{31}}, 556 (1962).
\item M.Harada, F.Sannino and J.Schechter, Phys.Rev. {\bf{D54}}, 1991 (1996);
      {\bf{D69}}, 034005 (2004), \
      M.Harada, {\it{PiPi scattering and scalar mesons in an effective chiral 
      lagrangian}}, hep-ph/0009051. 
\item M.Harada, S.Matsuzaki and K.Yamawaki, Phys.Rev. {\bf{D74}}, 076004 (2006). 
\item M.Bando,T.Kugo, S.Uehara, K.Yamawaki and T.Yanagida, 
      Phys.Rev.Lett. {\bf{54}}, 1215 (1985); \
      M.Harada, T.Kugo and K.Yamawaki, Prog.Thoer.Phys. {\bf{91}}, 801 (1994).

\item A.A.Tseytlin, Nucl.Phys. {\bf{B501}}, 41 (1997). 
\item S.Nagaoka, Prog.Theo.Phys. {\bf{110}}, 1219 (2004).
\item R.Parthasarathy and K.S.Viswanathan, ``$(\alpha')^4$ Corrections in Holographic 
Large $N_c$ QCD and $\pi-\pi$ Scattering'', arXiv: hep-th/0707512. 
\item O.Bergman, S.Seki and J.Sonnenschein, ``Quark mass and condensates in HQCD'',
      arXiv: hep-th/0708.2839; A.Dhar and P.Nag, ``Sakai-Sugimoto model, tachyon 
      condensation and chiral symmetry breaking'', arXiv: hep-th/0708.3233. 
\item R.Kami\'{n}ski, J.R.Pel\'{a}ez and F.J.Yndur\'{a}in, ``The pion-pion scattering 
amplitude. III: Improving the analysis with forward dispersion relations and Roy 
equations'', arXiv: hep-ph/0710.1150. 
\item D.Black, A.H.Fariborz, S.Moussa, S.Nasri and J.Schechter, Phys.Rev. {\bf{D64}}, 
014031 (2001). \ 
      D.Black and J.Gaunt, ``Study of scalar mesons in chiral Lagrangian frameworks'', 
      arXiv: hep-ph/0801.2472. 
\end{enumerate} 
\vspace{0.5cm} 

\begin{figure}
\begin{center}
\resizebox{120mm}{!}{\includegraphics{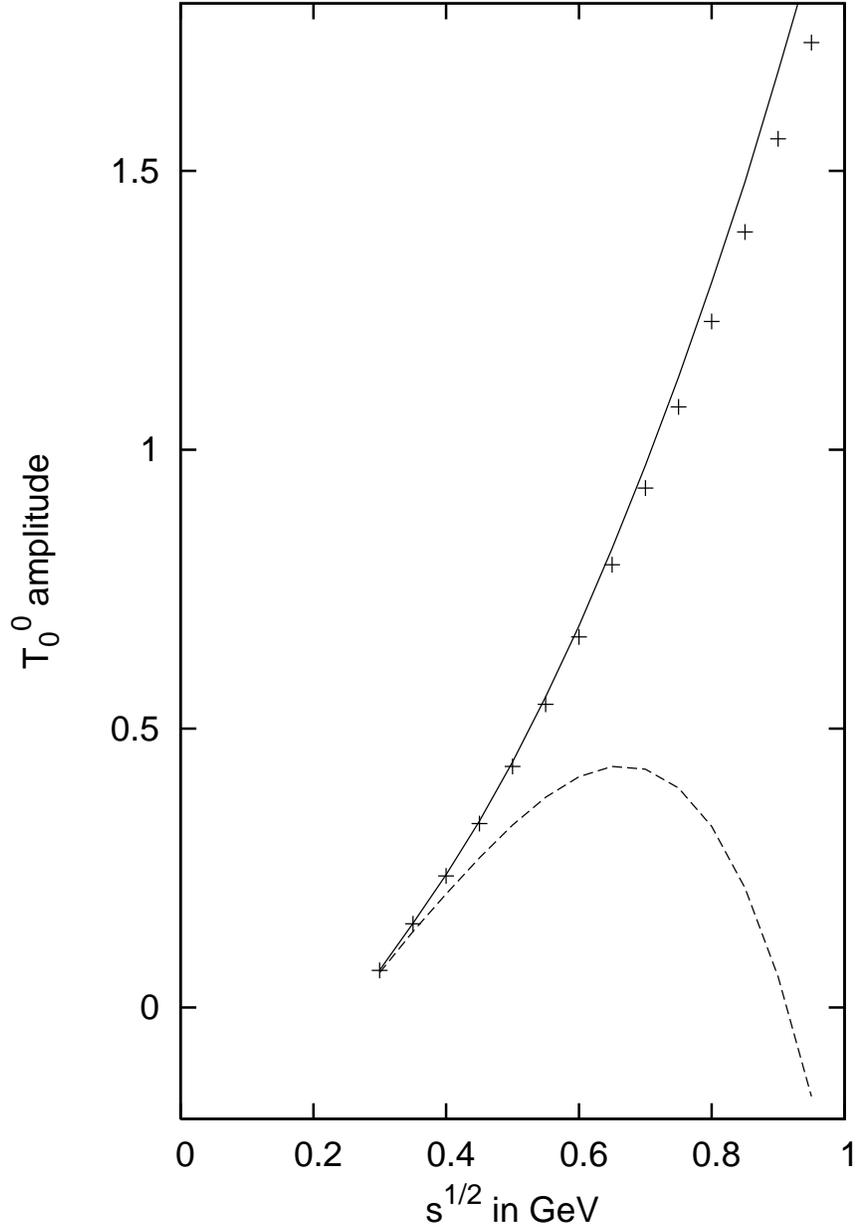}}
\caption{Continuous linearly rising curve is from (1). The points $+$ 
and the dotted curve are from 
(36) with $M_{KK}=0.94 GeV$ and $N_c=3$ and $N_c=11$ respectively.} 
\end{center}
\end{figure}

\vspace{0.5cm}

\begin{figure}
\begin{center}
\resizebox{120mm}{!}{\includegraphics{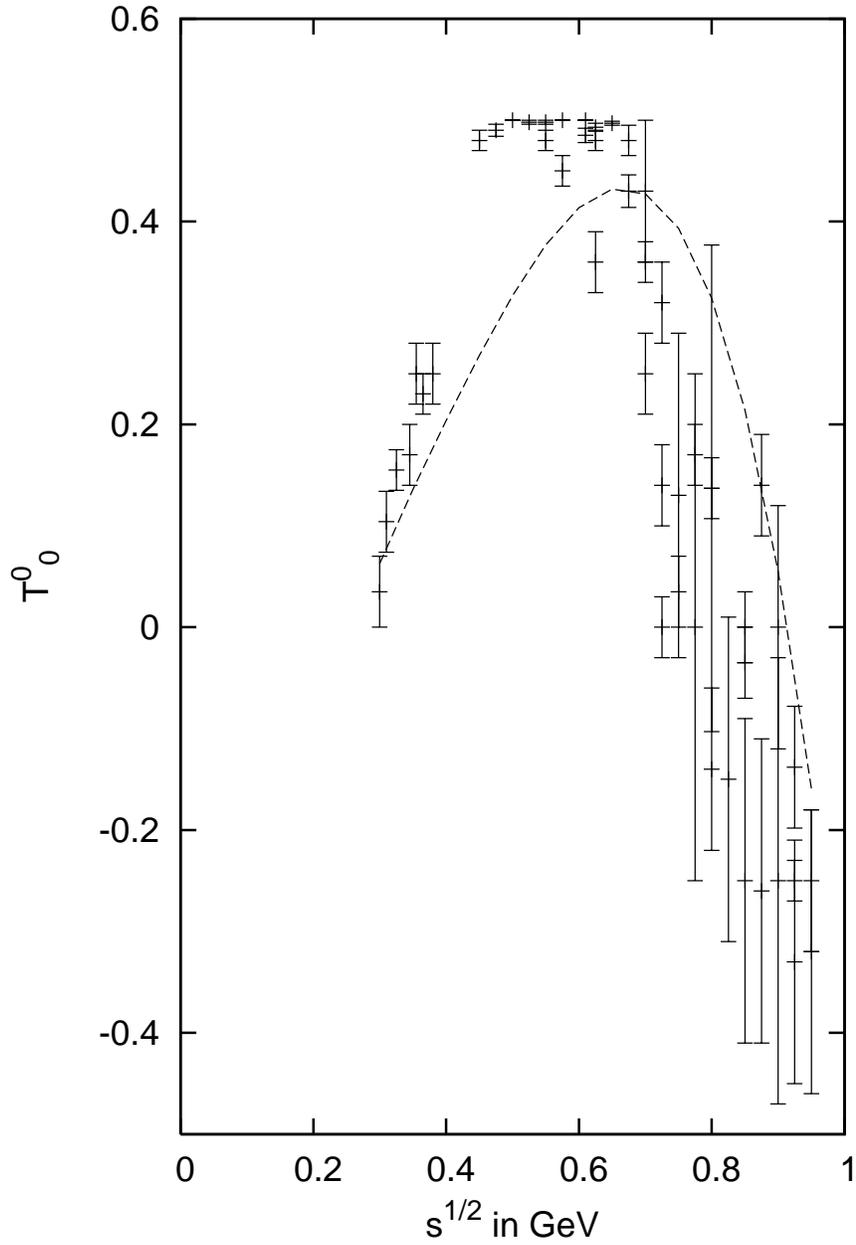}}
\caption{ $T^0_0$ amplitude from (36) (dotted curve) including ${\alpha'}^4$ corrections
and the  experimental data from [27]}
\end{center}
\end{figure}

\vspace{0.5cm}

\begin{figure}
\begin{center} 
\resizebox{120mm}{!}{\includegraphics{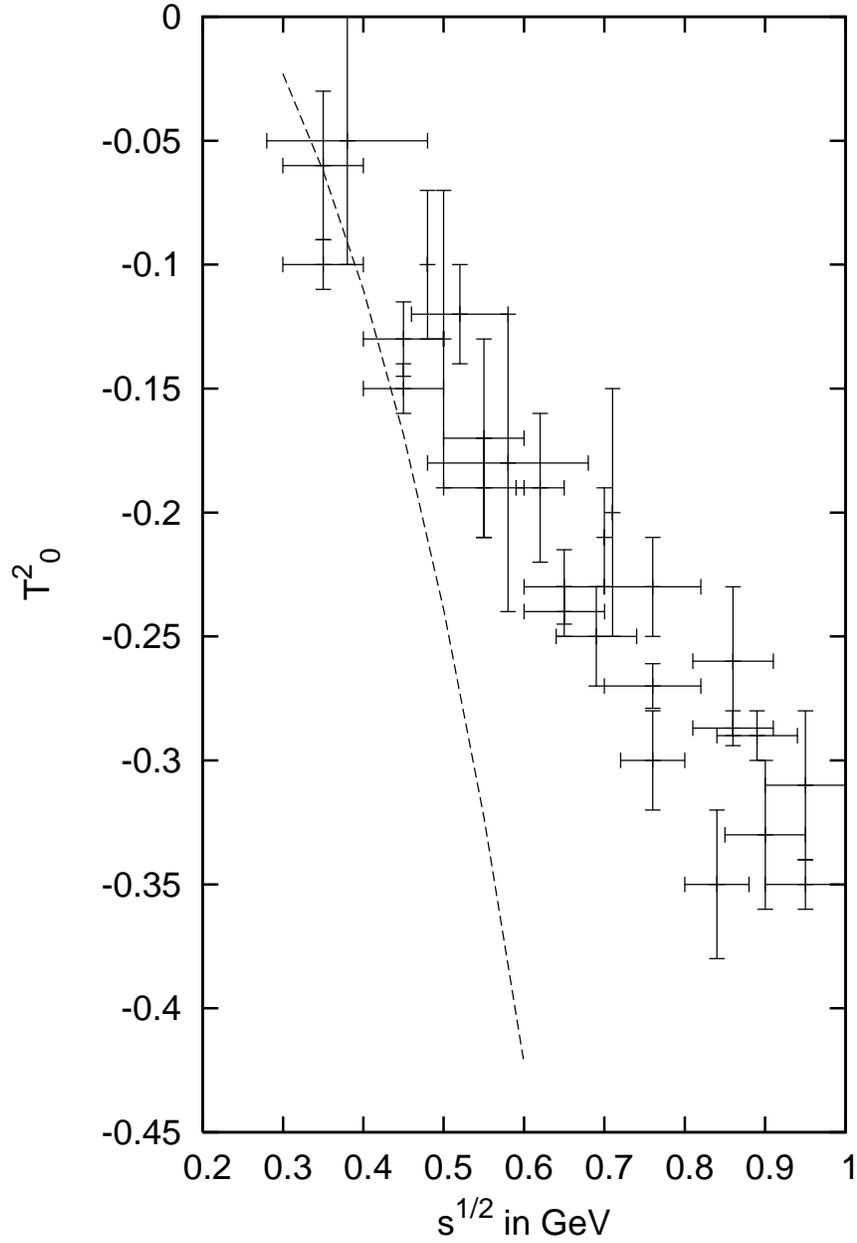}}
\caption{$T^2_0$ amplitude from (36) (dotted curve) including ${\alpha'}^4$ terms and 
the  experimental data from [27]} 
\end{center}
\end{figure}

\vspace{0.5cm} 

\begin{figure}
\begin{center}
\resizebox{120mm}{!}{\includegraphics{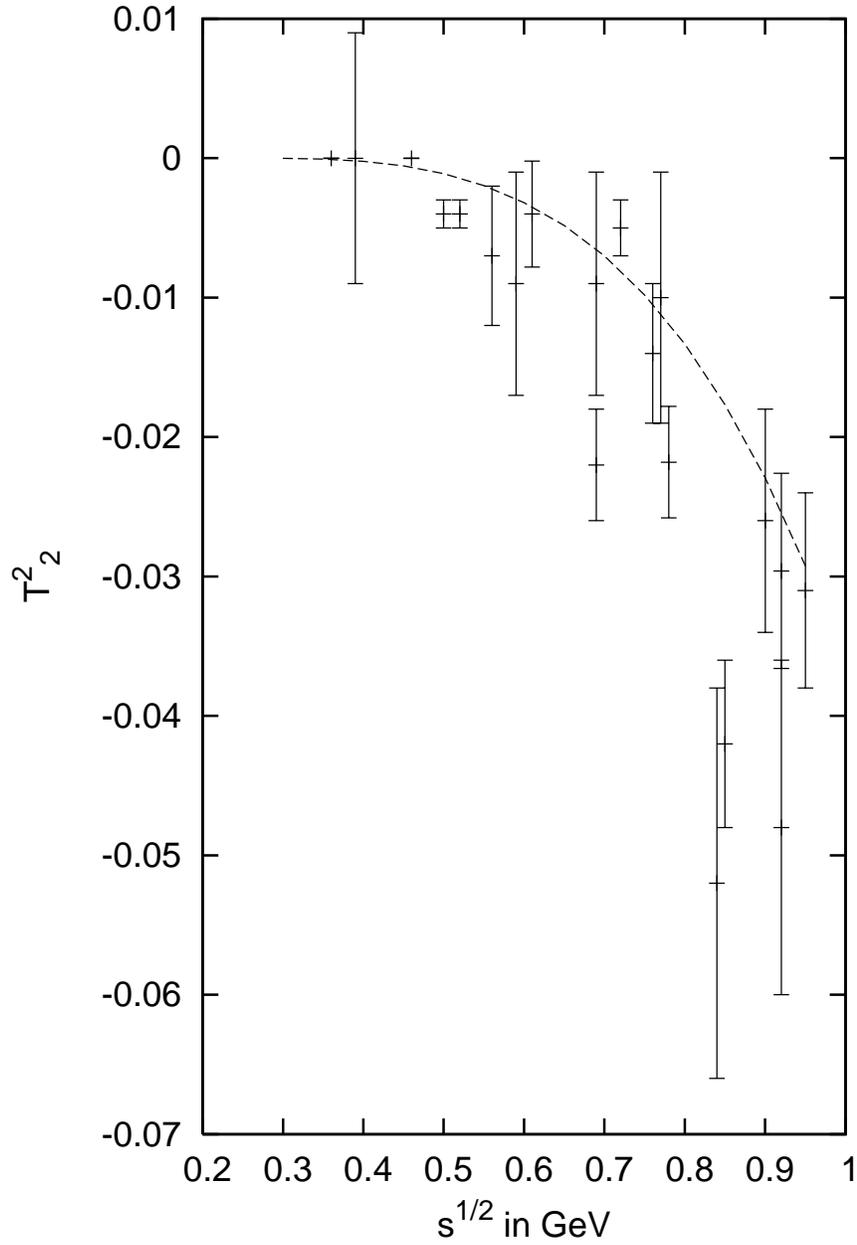}}
\caption{$T^2_2$ amplitude from (36) (dotted curve), only ${\alpha'}^4$ terms contribute
and the  experimental data from [27]} 
\end{center}
\end{figure}

\end{document}